\documentclass[aps,prl,twocolumn,superscriptaddress,showpacs]{revtex4}
\usepackage{graphicx}
\include{epsf}

\begin{document}

%\preprint{}

\title{A novel experimental approach for the detection of the dynamic 
Casimir effect}



\author{C. Braggio}
\affiliation{Dipartimento di Fisica dell'Universit\`a di Ferrara and INFN, Via 
del Paradiso 12, 44100 Ferrara, Italy}
\author{G. Bressi}
\affiliation{INFN, Sezione di Pavia, Via Bassi 6, 27100 Pavia, Italy}
\author{G. Carugno}
\affiliation{INFN, Sezione di Padova, Via Marzolo 8, 35131 Padova, Italy}
\author{C. Del Noce}
\email[]{delnoce@pd.infn.it}
\affiliation{INFN, Sezione di Padova, Via Marzolo 8, 35131 Padova, Italy}
%\email[]{delnoce@pd.infn.it}
\author{G. Galeazzi}
\affiliation{INFN - LNL, Viale dell'Universit\`a 2, 35020 Legnaro, Italy}
\author{A. Lombardi}
\affiliation{INFN - LNL, Viale dell'Universit\`a 2, 35020 Legnaro, Italy}
\author{A. Palmieri}
\affiliation{INFN - LNL, Viale dell'Universit\`a 2, 35020 Legnaro, Italy}
\author{G. Ruoso}
\affiliation{INFN - LNL, Viale dell'Universit\`a 2, 35020 Legnaro, Italy}
\author{D. Zanello}
\affiliation{INFN, Sezione di Roma, P.le A. Moro 2, 00185 Roma, Italy}
%\thanks{Contact Information: giuseppe.ruoso@lnl.infn.it}
\date{\today}

\begin{abstract} 
The Casimir effect is a well-known macroscopic consequence of 
quantum vacuum fluctuations, 
but whereas the static effect (Casimir force) has long
been observed experimentally, the dynamic Casimir effect  
is up to now undetected.
From an experimental viewpoint a possible detection
would imply the vibration of a mirror at gigahertz frequencies.
Mechanical motions at such frequencies turn out to be technically unfeasible.
Here we present a different experimental scheme where mechanical 
motions are avoided, and the results of laboratory tests showing that the scheme 
is practically feasible.
We think that at present this approach gives the only 
possibility of detecting this phenomenon.
\end{abstract}

%\maketitle
\pacs{12.20.Fv, 42.50.Dv, 13.40.--f}
%\keywords{Casimir force; Microresonators; Experimental test of 
%gravitational theories}
\maketitle

%\section{Introduction}

For \emph{any} quantum field, the \emph{vacuum} is defined as its ground 
state.
Differently than in the classic case, this ground state, due to the
uncertainty principle, is not empty, but 
filled with field fluctuations around a zero mean value.
Moreover this vacuum state depends on the field boundary conditions~:
if they change, there will be a correspondingly different vacuum (whose
fluctuations will have a different wavelength spectrum).
Thus a quantum vacuum state may be equivalent to real particles of a new vacuum
after a change in boundary conditions.

If we consider the \emph{electromagnetic field,} the peculiar nature of 
the quantum vacuum has experimentally observable consequences in the realm 
of microscopic physics, such as natural widths of spectral lines, Lamb 
shift, anomalous magnetic moment of the electron and many more.
It is perhaps even more striking that there exist also observable effects 
at a macroscopic level.
The Casimir force (static Casimir effect \cite{Cas, BMM}) is one of these 
macroscopic effects which has been observed experimentally.
A \emph{dynamic Casimir effect} is also predicted to occur when
one boundary is accelerated in a nonuniform way, as for instance when a 
metal surface undergoes harmonic oscillations.
In this case a number of virtual photons from the vacuum are converted
into real photons (``Casimir radiation''), while the moving metal surface 
loses energy \cite{Moo,FD, KG}.

It is worth notice that, whereas the static Casimir effect has been 
observed by several experiments \cite{Spa}, the Casimir radiation is to 
date unobserved, in spite of the abundant theoretical work done in this field 
\cite{LJR1, Dod, CDM, SSPS, LJR3}
(see \cite{Dod} for a historical review and a bibliography of the relevant studies).
We argue that this lack of experimental activity stems from the rooted
idea of using mechanical oscillations.
We shall show that this is unfeasible with 
present-day techniques.

Here we shall present a \emph{new experimental approach}
where an effective motion is generated by the 
excitation of a plasma in a semiconductor.
In terms of power this effective motion is much more convenient than a 
mechanical motion, since in a metal mirror only the conduction 
electrons reflect the electromagnetic waves, whereas a great amount of 
power would be wasted 
in the acceleration of the much heavier nuclei.
Some authors \cite{Dod2, CDLM, UPSS} have made use of our idea in order to 
construct 
a concrete model for theoretical calculations.
In this paper we wish to discuss the experimental details of 
our apparatus and to show the feasibility of a measurement.

%\section{Theoretical formulae}
\vspace*{\baselineskip}
We shall now \emph{outline the theoretical situation.}
The simplest system that can produce Casimir radiation is a single mirror, 
harmonically oscillating in a direction perpendicular to its surface.
In this case the number $N$ of created photons should be \cite{LJR3}~:
\begin{equation}
N = \frac{\omega t}{3 \pi}\left( \frac{v}{c} \right)^2 , 
\end{equation}
where $\omega$ is the angular
frequency of the mirror motion, $t$ is the duration of the motion, 
$v$ is the maximum speed reached in the oscillation, and $c$ the speed of 
light.
Even if we stretch all parameters to their utmost values
($\omega \sim 10^{10}$~rad/s, $t \sim 1$~s and $v/c \sim 10^{-8}$) the 
number of produced photons is not detectable.

A great theoretical progress was to realize that, 
when the oscillating mirror is a wall of an electromagnetic resonant 
cavity, the 
cavity itself behaves as a multiplier for the produced radiation, if the 
frequency of the moving wall is twice one of the proper electromagnetic 
cavity frequencies (parametric resonance).
It is however disappointing that the formulae developed so far
using different approaches (in the case of parametric resonance) are not 
the same and even irreconcilable.
Apart from minor differences the formulae for the produced photons 
found in literature \cite{LJR1, Dod, CDM, SSPS}
can be brought back to either of two forms~:
\begin{eqnarray}
N & = & \frac{\omega t}{2 \pi} \left( \frac{v}{c} \right)^2 Q , 
\label{photons} \\
N & = & \sinh^2 \left( \omega t \, \frac{v}{c} \, \right) , 
\label{exponential} 
\end{eqnarray}
where $Q$ is the quality factor of the cavity.
We shall show that with our proposed method even in 
the worst case (formula (\ref{photons})) the number of produced photons is 
sufficient for detection.

\vspace*{\baselineskip}
An \emph{experiment based on} the \emph{mechanical motion} 
of a resonant cavity wall would be \emph{too difficult for present-day 
techniques.}
The highest frequency attainable for mechanical motion is in the 
gigahertz range \cite{BD, YB} 
and following the parametric amplification request this implies microwave 
cavities with dimensions ranging from 1~cm to 1~m.
The motion of a single wall of such a cavity requires a huge amount of 
power.
In fact a wall of volume $V$ made of a material with mass density $\rho$ 
vibrating
at an angular frequency $\omega$ with an amplitude $\delta x$ has a
maximum kinetic energy $E = \frac{1}{2} \rho V \omega^2 \delta x^2$ which
vanishes in a time of order $\frac{ \pi}{2 \omega}$.
If we estimate the required power for $\rho = 3 \cdot 10^3~\mbox{kg/m}^3$,
$V=3 \, \mbox{cm} \times 3 \, \mbox{cm} \times 0.1 \, \mbox{mm} = 9 \cdot 
10^{-8}
\mbox{m}^3$, $\frac{\omega}{2\pi} = 2$~GHz, $\delta x=1$~nm, we obtain
about $3 \cdot 10^8$~W.

At present there are two known ways to make a body oscillate at gigahertz
frequencies and both of them have some disadvantages precluding their use 
in a dynamic Casimir experiment.
The first way would exploit acoustic waves in solids.
Waves at gigahertz frequencies were produced in the 60's by B\"ommel and
Dransfeld in a quartz rod placed inside a microwave resonant cavity
\cite{BD}. 
What makes this technique ineffective for our purpose is that a large
microwave power is needed and that the rod motion has a
maximum displacement $\delta x$ much less than 1~nm.
A small amplitude $\delta x$ implies a small maximum 
oscillation speed $v$
(for a harmonic motion $v = \omega \cdot \delta x$ where $\omega$ is the
oscillation angular frequency).
Hence the number of photons produced by a mechanical oscillation
with such a speed would be undetectable, as is readily seen from
formulae (\ref{photons}) and (\ref{exponential}).

The second technique is the one applied in acoustic microscopes \cite{YB}.
A resonant vibrating mode in a sapphire block is excited at a typical
frequency around 3~GHz.
The use of a mechanical system with high quality factor $Q$ reduces power
requests in this case.
For sapphire at 4.3~K the product of the cavity 
quality factor $Q$ by the
oscillation frequency $f$ is about $Q \cdot f \sim 10^{14}$~Hz \cite{BCT}.
Therefore if $f \sim 10^9$~Hz, $Q$ can be as high as $10^5$.
The same oscillation amplitude as in a nonresonant system can be reached
with a power $10^5$ times smaller.
But again the oscillation amplitude $\delta x$ is about $10^{-10}$~m and
the moved area is quite small (about 100~$\mu \mbox{m}^2$).

\vspace*{\baselineskip}
We now present our \emph{idea for} realizing \emph{an oscillating mirror} without 
employing mechanical methods.
The notion of using laser pulses to quickly change the dielectric properties 
of a semiconductor can be found in literature.
In 1989 Yablonovitch \cite{Yab} proposed the use of laser 
pulses to change the refraction index of a semiconductor very rapidly.
Another work by Lozovik, Tsvetus and Vinograd 
\cite{LTV} studied the parametric
excitation of electromagnetic waves using a dense plasma layer in a
cavity; the layer was created by irradiating a semiconductor film with
femtosecond laser pulses.

\begin{figure}
\centering
\includegraphics[height=6cm]{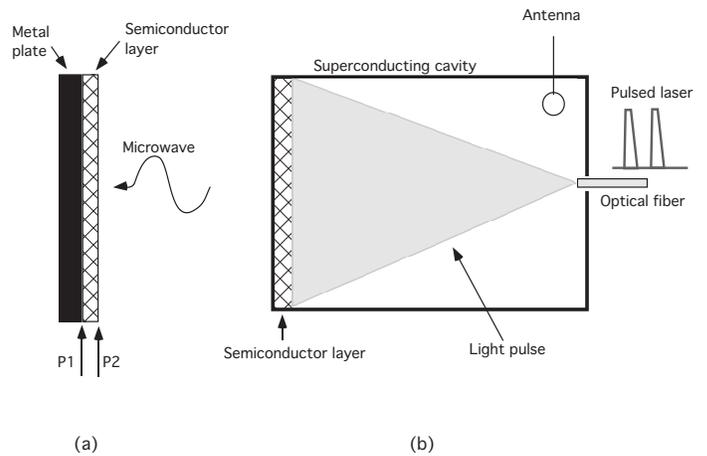}
\caption{
(a) Mirror effective motion~: a composite mirror changes its 
reflection
properties (under intermittent laser light irradiation), and the microwave 
reflecting surface switches its position between 
P1 and P2 accordingly.
(b) Arrangement of the composite mirror in a microwave resonant cavity.
The semiconductor is irradiated by an optical fiber piercing the cavity.}
\label{cavity}
\end{figure}
In \emph{our experimental scheme} we shall simulate a mirror motion by changing
the actively reflecting surface of a composite mirror.
The mirror consists of a metal plate with a semiconductor wafer
fixed on one side (see FIG.\ \ref{cavity}(a)).
The semiconductor reflectivity is driven by irradiation from laser light,
with photon energy corresponding to the semiconductor energy gap, so that 
it
can switch from completely transparent to completely reflective for
microwaves. 
By sending a train of laser pulses at a given frequency we get a mirror 
oscillating from position P1 to position P2.
An advantage of this method is that the distance between P1 and P2 can be made 
of the order of a millimeter, compared to about 1~nm obtainable 
by mechanical oscillations.
This leads to a layout as represented in FIG. 1(b).
The composite mirror becomes a wall of a superconducting cavity.
The laser pulses are guided into the cavity via an optical fiber.
A small pickup antenna is also introduced in the 
cavity 
and the signal fed to high sensitivity electronics.

However a number of \emph{points} need \emph{to be checked} in order that we may
state that this method is effective~:
\begin{enumerate}
\item[(1)] Is the mirror created in P2 as good as the one in P1~?
\item[(2)] Is it actually feasible to make the mirror appear and disappear in 
P2 at gigahertz frequencies~?
\item[(3)] Is the $Q$ of the cavity influenced by the presence of the 
semiconductor~?
\item[(4)] Is the sensitivity of the pickup electronics good enough to 
detect the predicted number of created photons~?
\end{enumerate}

The \emph{answers to these questions} have required about a year of laboratory 
tests \cite{BBCLPRZ, prep}.

\emph{Question} (1) was tackled by inserting a semiconductor layer in a waveguide
and measuring the reflected and the transmitted power under laser
irradiation.
It was proven that the semiconductor can reflect microwaves as effectively as 
copper.
This test yields also another important parameter, that is the laser power 
needed to make a good mirror.
This question arises from the fact that one needs to build a plasma of 
thickness equal to at least three 
skin depths (for the given microwave frequency) in order that it may be 
fully reflective.
The energy needed is 1~$\mu$J/cm$^2$ per pulse in the microwave 
range \cite{BBCLPRZ}.

The answer to \emph{question} (2) can be found in literature \cite{MSABL}.
The mirror appearance at P1 is fast enough for gigahertz frequencies,
since the transition time
of the electrons is some femtoseconds, so that the dominant factor is the
rise time of the laser pulse, which is in the hands of the experimenter.
However the disappearance of the mirror depends on the recombination time of
the electrons, which is a property of the semiconductor only.
If one uses semi-intrinsic semiconductors one can obtain recombination times 
as low as 5--10~ps \cite{MSABL}.

The answer to \emph{question} (3) was straightforward.
We measured the $Q$ value of a niobium cavity brought to 4.6~K, determining the 
decay time of the loaded cavity, and got $Q = 2 \cdot 10^6$.
Once the semiconductor wafer was inserted in the cavity no difference in the 
decay time (hence in $Q$) was detected.

In order to answer \emph{question} (4) a complete electronic chain 
was connected to the pickup antenna inserted in the cryogenic cavity.
The first amplification stage was placed near the cavity at liquid helium 
temperature \cite{bradley}.
The cavity was then loaded with microwave pulses of decreasing power in order 
that the minimum detectable signal might be reached.
The minimum signal detected had an energy of 0.1~eV, 
corresponding to about $10^4$ microwave 
(2.5~GHz) photons.
By taking 100 measurements one arrives at $10^3$ photons.
We think that further improvements in the electronic chain will allow us to 
detect even feebler signals.
A by-product of this measurement regards the possible source of noise due to 
thermal radiation from the electron-hole 
plasma created in the semiconductor when the laser shines.
We tried to measure this noise but it was below our sensitivity (in our 
frequency band).

\begin{figure}
\centering
\includegraphics[height=6cm]{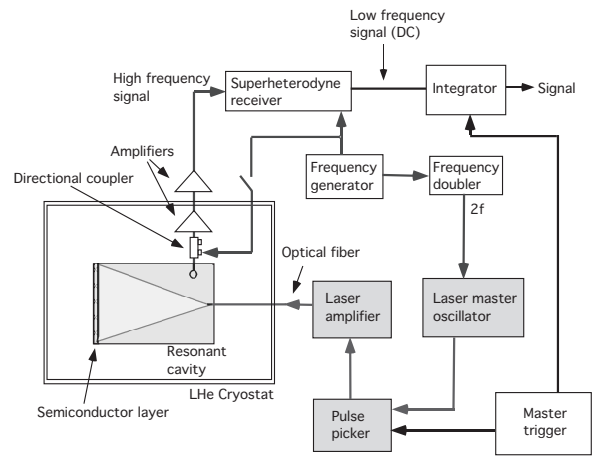}
\caption{Detailed experimental setup. 
There are three main parts~: the 
electromagnetic cavity already shown in FIG.~1, the electronic chain and the 
laser system. 
This block diagram displays the interrelations between laser and radio frequency 
generator for the control of parametric resonance.}
\label{detailed}
\end{figure}

\vspace*{\baselineskip}
On the basis of our previous results a general \emph{layout for the
detection of Casimir radiation} is shown in FIG.~\ref{detailed}.
A niobium cavity at cryogenic temperature is placed in a vacuum vessel.
A cryogenic amplifier \cite{bradley} is connected by a 
transmission line to an inductive pickup loop coupled with the cavity in 
critical matching.
A directional coupler is inserted between the cavity and the cryogenic 
amplifier to enable measurements of the resonance cavity reflection 
coefficient and calibration of the electronic chain.
The signal output by the cryogenic amplifier is further amplified at room 
temperature, then processed by a superheterodyne receiver and 
eventually integrated over time.
The laser light carried by the optical fiber is tuned in the near infrared
and modulated in amplitude at a frequency exactly double the cavity
resonance frequency.
The generator drives a frequency doubler whose output turns to a
low power laser master oscillator.
The master oscillator yields a continuous signal from which a pulse 
picker selects the number of pulses required in each excitation stage.
The total energy stored in the laser is limited, so must be the number of 
available pulses.
Our present estimate is
between $10^3$ and $10^4$ pulses for each run.

This experimental setup leaves open the possibility of changing many 
configuration parameters to help distinguishing real from spurious 
signals.
We can change the master laser frequency and thus the oscillation mirror 
frequency to slightly detune the parametric resonance condition.
Also we can vary the cavity temperature to study possible contributions 
from thermal radiation.
Mirrors made with different semiconductor samples and with different 
thickness can be tried.

\vspace*{\baselineskip}

In order to show that our scheme leads to observable results we
need to insert real numbers in the theoretical formulae 
and \emph{compare the 
predicted number of photons with our sensitivity.}
Several physical parameters are essentially already chosen, since a niobium 
cavity and an electronic chain have been used satisfactorily in the tests 
carried on to answer questions (3) and (4).
The niobium cavity has transverse dimensions of 71~mm and 22~mm, and 
length $x = 110$~mm. 
The cavity mode chosen was TE$_{101}$ with eigenfrequency around 2.5~GHz.
The semiconductor was GaAs with thickness $2\, \delta x = 0.6$~mm.
The excitation time duration for a single run, at 5~GHz, according to the 
number of pulses, can be between 0.2 and 2~$\mu$s.
Typically a run can be repeated after a few seconds.
The following data can be used to estimate the number of photons 
produced by dynamic Casimir effect~:
\begin{eqnarray*}
\frac{\omega}{2\pi} &=& 2.5 \cdot 10^9 \mbox{s}^{-1} \, ; \qquad \qquad 
\qquad
t = 10^{-6} \mbox{s} \, ; \nonumber \\ 
\frac{v}{c} = \frac{\delta x}{x} &=& 
\frac{0.3\mbox{~mm}}{110\mbox{~mm}} = 3\cdot 10^{-3} \, ; \qquad Q 
= 2 \cdot 
10^6 \, . \nonumber
\end{eqnarray*}
With formula (\ref{photons}), which is the more pessimistic, we find that this
number is $4 \cdot 10^4$, well above our sensitivity.

\vspace*{\baselineskip}
The above figures foster our hopes that eventually some light will be shed 
on pressing open problems.
In fact a good knowledge of quantum vacuum is of great \emph{importance in 
cosmology,} 
both to the recurrent question of Einstein's cosmological constant \cite{Wei},
with its significance to the dark matter problem; and to the critical 
question of 
the birth of 
density inhomogeneities, ancestors of galaxies, from inflated quantum vacuum fluctuations 
\cite{Dav}.
Moreover a sound grasp of quantum vacuum dynamics is crucial in understanding some
issues on the nature of quantum particles and on the relationships among vacuum
noise, the concepts of information and entropy, and gravitation \cite{Dav}.

\vspace*{\baselineskip}
We wish to thank L. Badan for technical help with cryogenics, 
E. Berto for mechanical support, and D. Corti for help with radio 
frequency measurements.

\end{document}